\pdfoutput=1
\newcommand*{\ATLASLATEXPATH}{atlaslatex/latex/}
\documentclass[cernpreprint,txfonts,UKenglish,texlive=2016]{\ATLASLATEXPATH atlasdoc}
\usepackage{\ATLASLATEXPATH atlaspackage}
\usepackage{\ATLASLATEXPATH atlasbiblatex}
\usepackage{\ATLASLATEXPATH atlasphysics}
\usepackage{\ATLASLATEXPATH atlasprocess}
\graphicspath{{atlaslatex/logos/}{figures/}}
\usepackage{paper-defs}
\usepackage{multirow}
\usepackage{bigdelim}
\AtlasTitle{Measurement of the $t\bar{t}\gamma$ production cross section in
proton--proton collisions at $\sqrt{s} = 8$ Te\kern -0.1em V with the ATLAS
detector}

\author{The ATLAS Collaboration}

\PreprintIdNumber{CERN-EP-2017-086}
\AtlasJournalRef{JHEP 11 (2017) 086}
\AtlasDOI{10.1007/JHEP11(2017)086}

\AtlasAbstract{The cross section of a top-quark pair produced in association
with a photon is measured in proton--proton collisions at a centre-of-mass
energy of $\sqrt{s} = 8$ Te\kern -0.1em V with $20.2$ fb$^{-1}$ of data
collected by the ATLAS detector at the Large Hadron Collider in 2012.  The
measurement is performed by selecting events that contain a photon with
transverse momentum $p_\mathrm{T} > 15\;\mathrm{Ge\kern -0.1em V}$, an isolated
lepton with large transverse momentum, large missing transverse momentum, and
at least four jets, where at least one is identified as originating from a
$b$-quark.  The production cross section is measured in a fiducial region close
to the selection requirements.  It is found to be $139 \pm
7\,(\mathrm{stat.\!}\,) \pm 17\,(\mathrm{syst.\!}\,)\,\mathrm{fb}$, in good
agreement with the theoretical prediction at next-to-leading order of $151 \pm
24$ fb.  In addition, differential cross sections in the fiducial region are
measured as a function of the transverse momentum and pseudorapidity of the
photon.}

\hypersetup{pdftitle={ATLAS document},pdfauthor={The ATLAS Collaboration}}

\begin{document}
\maketitle
\tableofcontents
\newpage
\section{Introduction} 
\label{s:intro}

Measurements of top-quark properties play an important role in testing the
Standard Model (SM) and its possible extensions.  Studies of the production and
dynamics of a top-quark pair in association with a photon (\ttg) probe the
$t\gamma$ electroweak coupling. For instance, deviations in the \pT spectrum of
the photon from the SM prediction could point to new physics through anomalous
dipole moments of the top quark, as discussed in refs.~\cite{Baur:2004uw,
Bouzas:2012av, Rontsch:2015una, Schulze:2016qas, Bylund:2016phk, Duan:2016qlc}.  

Photons can originate not only from top quarks, but also from their decay
products, including the quarks and leptons from the decay of the $W$ bosons.
In addition, they can be radiated from incoming partons.  Evidence for the
production of a top-quark pair in association with an energetic, isolated
photon has been found in \ppbar collisions at the Tevatron collider at a
centre-of-mass energy of $\sqrt{s} = \SI{1.96}{\TeV}$ by the CDF
Collaboration~\cite{Aaltonen:2011sp}.  Finally, observation of the \ttg process
was reported by the ATLAS Collaboration in $pp$ collisions at $\sqrt{s}
=\SI{7}{\TeV}$~\cite{TOPQ-2012-07}.

This paper describes a measurement of the \ttg production cross section, based
on a data set recorded with the ATLAS detector in 2012 at a centre-of-mass
energy of $\sqrt{s} = \SI{8}{\TeV}$ and corresponding to an integrated
luminosity of \SI{20.2}{\ifb}.  The cross section is measured with a
maximum-likelihood fit using templates defined within a fiducial volume chosen
to be close to the selection requirements implemented in the analysis.  Only
final states with exactly one reconstructed lepton (electron or muon),
including those originating from $\tau$ lepton decays, are considered.  These
final states are referred to as the single-lepton channel in the following.  In
addition to the inclusive cross section, differential cross sections as a
function of the transverse momentum \pt and the pseudorapidity $\eta$ of the
photon are measured for the same fiducial volume.  The cross sections are
compared to the theoretical calculations at next-to-leading order
(NLO)~\cite{melnikov} in the strong interaction.

\section{ATLAS detector}
\label{s:atlas}

The ATLAS detector is described in detail elsewhere~\cite{PERF-2007-01}. Here,
a short overview is presented with a focus on the electromagnetic calorimeter,
which provides an accurate measurement of energetic photons.  The major
components of the ATLAS detector are an inner detector (ID) surrounded by a
thin superconducting solenoid providing a \SI{2}{T} axial magnetic field,
electromagnetic (EM) and hadronic calorimeters, and a muon spectrometer (MS).
The ID provides tracking information and is composed of three subsystems.  The
pixel and silicon microstrip detectors cover the pseudorapidity range $\abseta
< 2.5$,\footnote{ATLAS uses a right-handed coordinate system with its origin at
the nominal interaction point (IP) in the centre of the detector and the
$z$-axis along the beam pipe. The $x$-axis points from the IP to the centre of
the LHC ring, and the $y$-axis points upward. Cylindrical coordinates
$(r,\phi)$ are used in the transverse plane, $\phi$ being the azimuthal angle
around the $z$-axis. The pseudorapidity is defined in terms of the polar angle
$\theta$ as $\eta=-\ln\tan(\theta/2)$.} while the transition radiation tracker
has an acceptance range of $\abseta < 2.0$ and provides identification
information for electrons.  The MS consists of a large superconducting air-core
toroidal magnet system, three stations of chambers for high-precision tracking
measurements in the region $\abseta < 2.7$, and a muon trigger system effective
over the region $\abseta < 2.4$.

The electromagnetic calorimeter (EMC) is a lead/liquid-argon detector composed
of a barrel ($\abseta < 1.475$) and two endcaps ($1.375 < \abseta < 3.2$). For
$\abseta < 2.5$, the calorimeter has three layers, longitudinal in shower
depth, with the first layer having the highest granularity in the $\eta$
direction, and the second layer collecting most of the electromagnetic shower
energy for high-\pT electrons or photons.  A thin presampling layer in the
range $\abseta < 1.8$ is used to correct for the energy lost by EM particles
upstream of the calorimeter.  The hadronic calorimeter system, which surrounds
the electromagnetic calorimeter, is based on two different active media,
scintillator tiles or liquid argon, and with steel, copper, or tungsten as the
absorber materials.  Photons are identified as narrow, isolated showers in the
EMC with no spill-over into the hadronic calorimeter. The fine segmentation of
the ATLAS calorimeter allows efficient rejection of jets fragmenting to
high-energy $\pi^{0}$ or $\eta$ mesons that could be misidentified as isolated
prompt photons.

\section{Data and simulation samples}
\label{s:samples}

Monte Carlo simulated events are used to evaluate signal efficiencies and
backgrounds, and to estimate and correct for resolution effects.  Additional
simulated inelastic $pp$ collisions, generated with
\PYTHIA~8.160~\cite{PythiaManual8} using the \pdf{MSTW2008} LO
\cite{Martin:2009iq} parton distribution functions (PDFs) and parameter values
set according to the \tune{A2} tune~\cite{ATL-PHYS-PUB-2012-003}, are overlaid
to simulate the effects of pile-up from additional interactions in the same and
nearby bunch crossings. 

The \ttg signal sample is modelled with \MGMCatNLO~2.1.0~\cite{Alwall:2014hca},
interfaced  to \PYTHIA 6.427~\cite{PythiaManual} for parton showering and
hadronisation.  The \ttg matrix element is generated at leading order (LO)
including the decays of the top quarks.  The photons can be emitted from the
incoming partons, from the top quark, or from the decay products.  The
renormalisation (\muren) and factorisation (\mufac) scales are chosen such that
the top-quark pair (\ttbar) events, produced with the same settings as used for
the \ttg production, describe the \ttbar data well. The scale choice is made to
have the best agreement with the results in ref.~\cite{TOPQ-2011-21}.  Good
agreement with the data is obtained for $\muren=\mufac=2\mtop$, where
$\mtop=\SI{172.5}{\GeV}$ is the mass of the top quark used in all simulated
samples.  In order to avoid infrared and collinear singularities, the following
kinematic requirements are applied to the generated events:  the photon
transverse momentum must be larger than \SI{10}{\GeV} and the absolute value of
its pseudorapidity less than 5, at least one charged lepton with a transverse
momentum larger than \SI{15}{\GeV}, and the distance $\dR = \sqrt{(\dEta)^2 +
(\dPhi)^2}$ between the photon and all other charged particles in the final
state must be larger than $0.2$.  The \pdf{CTEQ6L1} (LO) PDF~\cite{cteq6} is
used both for the matrix element and in the \PYTHIA~6 parton shower, using the
corresponding \tune{Perugia2011C} tune~\cite{Skands:2010ak}.  Additional photon
radiation is generated with \textsc{Photos}~2.15.4~\cite{Barberio:1993qi,
Golonka:2005pn}.  The decays of the $\tau$ leptons are handled by
\textsc{Tauola}~2.7~\cite{Golonka:2003xt}.  The signal sample is simulated for
single-lepton and dilepton \ttg events.  The sample is normalised to the NLO
prediction using $K$-factors, as described in section~\ref{s:fit}. 

The production of $W\gamma +$jets and $Z\gamma +$jets final states is simulated
with the \SHERPA~1.4.0~\cite{sherpa} event generator, using the NLO \pdf{CT10}
PDF~\cite{Lai:2010vv}. The matrix elements are calculated at LO with up to
three final-state partons. The parton shower uses \SHERPA's default set of
tuned parameters.  To evaluate the uncertainty due to $W\gamma +$jets
modelling, an additional sample generated with \ALPGEN~2.14~\cite{Alpgen},
interfaced to \PYTHIA~6.426, is used. For both the matrix-element calculations
and the parton-shower evolution, the \pdf{CTEQ6L1} PDF is used.  The tuned
parameters for the parton shower, the additional photon radiation, and $\tau$
lepton decays are handled in the same way as for the signal \ttg sample.  No
dedicated samples were produced for $Z\gamma +$jets final states due to the
expected small contribution to the total uncertainty.

Electroweak production of the top quark (single-top) is simulated with
\POWHEGBOX~v1.0~\cite{Alioli:2009je, Re:2010bp} using the NLO \pdf{CT10} PDF.
The same models are employed for parton shower, photon radiation, and $\tau$
lepton decays as used for the signal \ttg sample.  The event generator versions
for these processes are the ones used for the \ttg sample, apart from
\PYTHIA~6.426 being used for the showering of the $Wt$ channel.  In order to
remove the overlap with \ttbar, the $Wt$ sample is produced using the so-called
diagram removal generation scheme~\cite{Frixione:2008yi}.  The $t$-channel and
$Wt$-channel samples are normalised to the approximate next-to-next-to-leading
order calculations in refs.~\cite{Kidonakis:2011wy}
and~\cite{Kidonakis:2010ux}, respectively.  In addition, diboson ($WW$, $WZ$,
$ZZ$) production samples are generated using \ALPGEN~2.14~\cite{Alpgen},
interfaced to \HERWIG~6.520~\cite{herwig} and \Jimmy~4.31~\cite{jimmy}.
For these, the \pdf{CTEQ6L1} PDF and the corresponding \tune{AUET2}
tune~\cite{ATL-PHYS-PUB-2011-008} are used.  The additional photon radiation
and $\tau$ lepton decays are handled in the same way as for the signal \ttg
sample.  The samples are normalised using the NLO calculation in
ref.~\cite{Campbell:2011bn}.

The \ttbar process is simulated using the NLO quantum-chromodynamics (QCD)
matrix-element event generator \POWHEGBOX~v1.0 \cite{Powheg, Powbox1, Powbox2},
using the \pdf{CT10} PDF and interfaced to \PYTHIA~6.427 for the parton shower,
fragmentation and underlying-event modelling. Other settings are handled in the
same way as for the signal \ttg sample. This sample is used exclusively to
validate the data-driven backgrounds.   

All samples are processed either through the full ATLAS detector
simulation~\cite{SOFT-2010-01} based on \GEANT 4~\cite{Agostinelli:2002hh}, or
through a faster simulation using parameterised showers in the
calorimeters~\cite{ATL-PHYS-PUB-2010-013}.  The resulting simulated events are
processed with the same reconstruction algorithms and analysis chain as the
data.

The measurement is based on data collected by the ATLAS experiment in $pp$
collisions at $\sqrt{s} = \SI{8}{\TeV}$ in 2012.  The corresponding integrated
luminosity is \SI{20.2}{\ifb}.  The absolute luminosity scale is derived from
beam-separation scans performed in November 2012~\cite{DAPR-2013-01}. 

\section{Event reconstruction and selection}
\label{s:selection}

The single-lepton \ttg final state is characterised by the presence of a
high-\pt photon, an isolated lepton with a large \pt, large missing transverse
momentum originating from the neutrino in the leptonic decay of a $W$ boson,
two jets from the hadronic decay of the other $W$ boson, and two $b$-quark
jets.

\subsection{Event reconstruction}
\label{s:objects}

Electron candidates are reconstructed from energy deposits in the central
region of the EMC with associated tracks in the ID~\cite{PERF-2013-03}. The
candidates have to satisfy the tight identification
criteria~\cite{Aaboud:2016vfy} and  are required to have a transverse momentum
of $\pT > \SI{25}{\GeV}$ and $|\etacl|<2.47$, excluding the transition region
between barrel and endcap EM calorimeters, \mbox{$1.37 < |\etacl| < 1.52$}.
The variable $\etacl$ is the pseudorapidity of the electromagnetic energy
cluster with respect to the geometric centre of the detector.  Electrons are
further required to have a longitudinal impact parameter with respect to the
primary vertex\footnote{Events are required to contain a hard collision
primary vertex with at least four associated charged particle tracks of $\pT >
\SI{0.4}{\GeV}$.  If there are multiple primary vertices in an event, the one
with the largest sum of track $\pT^2$ is selected.} of less than \SI{2}{mm}.
Isolation requirements on calorimeter and tracking variables are used to reduce
the background from jets misidentified as electrons.  The calorimeter isolation
variable is based on the energy sum of cells within a cone of size $\dR = 0.2$
around the direction of each electron candidate. This energy sum excludes cells
associated with the electron's cluster and is corrected for leakage from the
electron's cluster itself and for energy deposits from pile-up. The tracking
isolation variable is based on the sum of the transverse momenta of all tracks
around the electron in a cone of size $\dR = 0.3$, excluding the electron
track.  For both isolation variables, requirements are chosen to give
separately a $90\%$ electron selection efficiency for electrons from \Zee
decays in each \pT bin.

Muon candidates are identified by matching tracks in the muon spectrometer with
tracks in the ID and are required to have $\abseta < 2.5$ and $\pT >
\SI{25}{\GeV}$~\cite{PERF-2011-01, PERF-2014-05}.  The longitudinal impact
parameter with respect to the primary vertex is required to be less than
\SI{2}{mm}.  To reduce the background from muons originating from heavy-flavour
decays inside jets, muons are required to be separated by $\dR > 0.4$ from the
nearest jet and to be isolated.  The muon isolation variable, defined as the
ratio of the sum of \pT of tracks, excluding the muon, in a cone of variable
size $\dR = \SI{10}{\GeV} / \pT(\mu)$, to the \pT of the muon is required to be
smaller than $0.05$~\cite{Rehermann:2010vq}.  This isolation requirement
results in an efficiency of about $97\%$ for muons from \Zmm decays.

Reconstruction of photons~\cite{PERF-2013-04} starts with forming seed clusters
in the EMC with a size $0.075 \times 0.123$ in the $\eta \times \phi$ space and
with transverse energy above \SI{2.5}{\GeV} using the sliding-window
algorithm~\cite{ATL-LARG-PUB-2008-002}.  Tracks in the ID are matched to EM
seed clusters and are used to form conversion-vertex candidates where possible.
The  conversion-vertex candidates are then matched to the seed clusters.  A
final algorithm decides whether a seed cluster corresponds to an unconverted
photon, a converted photon or a single electron based on the matching to
conversion vertices or tracks and on the cluster and track(s) four-momenta.
Photon candidates must fulfil identification criteria based on shower-shape
variables containing information from the first layer of the EMC (strip
layer)~\cite{PERF-2013-05}.  All photons are required to have a transverse
momentum of at least \SI{15}{\GeV} and $|\etacl|<2.37$, excluding the
transition region of the EMC, $1.37 < |\etacl| < 1.52$. 

Jet candidates are reconstructed with the anti-$k_t$~\cite{Cacciari:2008gp}
algorithm with a radius parameter $R = 0.4$, using cell
clusters~\cite{PERF-2014-07} in the calorimeter, which are calibrated using the
local cluster weighting method~\cite{PERF-2011-03}.  The energies of jets are
then calibrated using an energy- and $\eta$-dependent simulation-based
calibration scheme with in situ corrections based on data.  Jets are required
to have $\pt> \SI{25}{\GeV}$ and $\abseta < 2.5$.  To suppress jets from
pile-up, a requirement on the jet vertex fraction (JVF) is imposed. It is
defined as the ratio of the sum of the \pT of tracks associated with both the
jet and the primary vertex to the sum of the \pT of all tracks associated with
the jet.  Jets with $\pT < \SI{50}{\GeV}$ and $\abseta < 2.4$ are required to
satisfy $|\text{JVF}| > 0.5$~\cite{PERF-2014-03}, which achieves high
efficiency for jets from hard scatters and high rejection of pile-up jets.

Muons within a cone of $\dR=0.4$ around a jet are removed, as well as the
closest jet within a cone of $\dR = 0.2$ (\num{0.1}) around an electron
(photon).  Finally,  electrons within a cone of $\dR =0.4$ around a jet are
removed.  

Jets containing $b$-hadrons are tagged by an algorithm that uses multivariate
techniques combining information from the impact parameters of displaced tracks
as well as topological properties of secondary and tertiary decay vertices
reconstructed within the jet~\cite{PERF-2012-04, ATLAS-CONF-2014-046,
ATLAS-CONF-2014-004}.  The working point used for this selection corresponds to
a 70\% efficiency for $b$-jets with $\pT > \SI{20}{\GeV}$ and $\abseta < 2.5$
in simulated \ttbar events, and a light-quark or gluon jet rejection factor of
$140$.

The missing transverse momentum, with magnitude \MET, is reconstructed as the
transverse component of the negative vector sum of the momenta of all
electrons, muons, photons and jets, as well as calibrated calorimeter energy
clusters not associated with any of the above~\cite{PERF-2014-04}.

\subsection{Event selection}
\label{s:evsel}

A single-lepton trigger is used to select the events. For the electron channel,
the trigger requires either a \pT threshold of \SI{24}{\GeV} and isolation, or
a \pT threshold of \SI{60}{\GeV} independent of isolation. For the muon
channel, the trigger requires a \pt threshold of \SI{24}{\GeV} and isolation,
or \SI{34}{\GeV} independent of isolation.

In both channels, the presence of exactly one lepton with $\pt > \SI{25}{\GeV}$
passing the trigger requirements and at least four jets is required. In order
to reduce several background contributions, but mainly the $W+$jets background,
at least one jet is required to be tagged as a $b$-jet. 

For the muon channel, additional requirements on the missing transverse
momentum and on \mtw, the transverse mass of the $W$ boson
candidate~\cite{TOPQ-2010-01}, $\MET > \SI{20} {\GeV}$ and $\MET + \mtw >
\SI{60}{\GeV}$, are imposed.  For the electron channel the requirements are
tighter, $\MET > \SI{30}{\GeV}$  and $\mtw> \SI{30}{\GeV}$, due to the larger
multijet background. 

The \ttg candidates are selected by applying the above criteria and by
requiring exactly one photon.  Events with a jet within a cone of $\dR=0.5$
around the selected photon are rejected to remove photon radiation from quarks.
In order to enrich the sample with events in which a photon is radiated from a
top quark, the distance between the selected photon and lepton direction is
required to be larger than $\dR = 0.7$. For the electron channel, the invariant
mass of the electron and the photon has to be outside a \SI{5}{\GeV} mass
window around the $Z$ boson mass (i.e. $m_{e\gamma} < \SI{86}{\GeV}$ or
$m_{e\gamma} > \SI{96}{\GeV}$) in order to suppress $Z$+jet events with one
electron misidentified as a photon.  Additional control regions are defined in
section~\ref{s:templates}.

\begin{figure}[htbp]
\centering
{
  \subfloat[]{\includegraphics[width=0.49\textwidth]{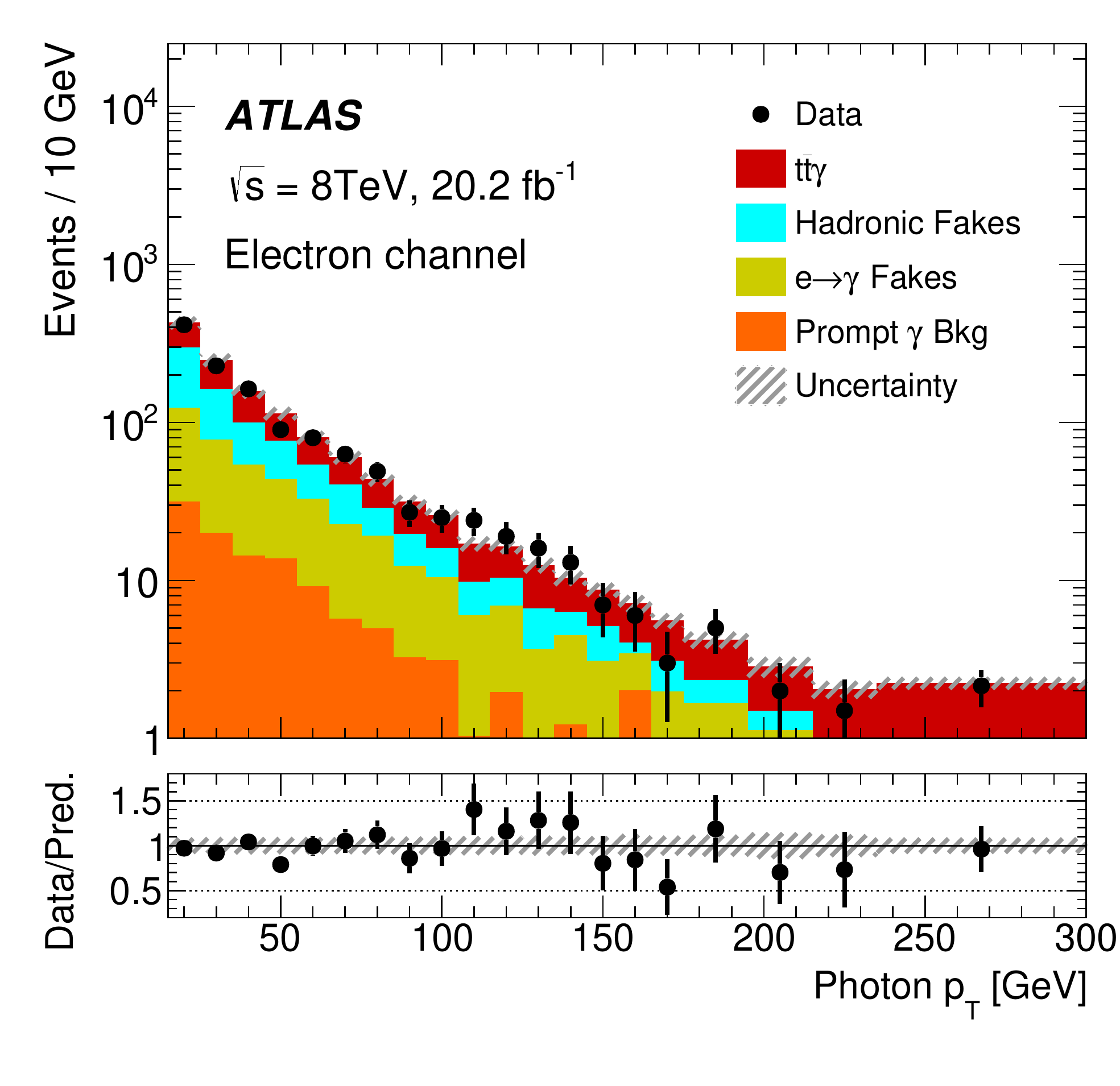}\label{fig:sel_el_ph_pt}}
  \subfloat[]{\includegraphics[width=0.49\textwidth]{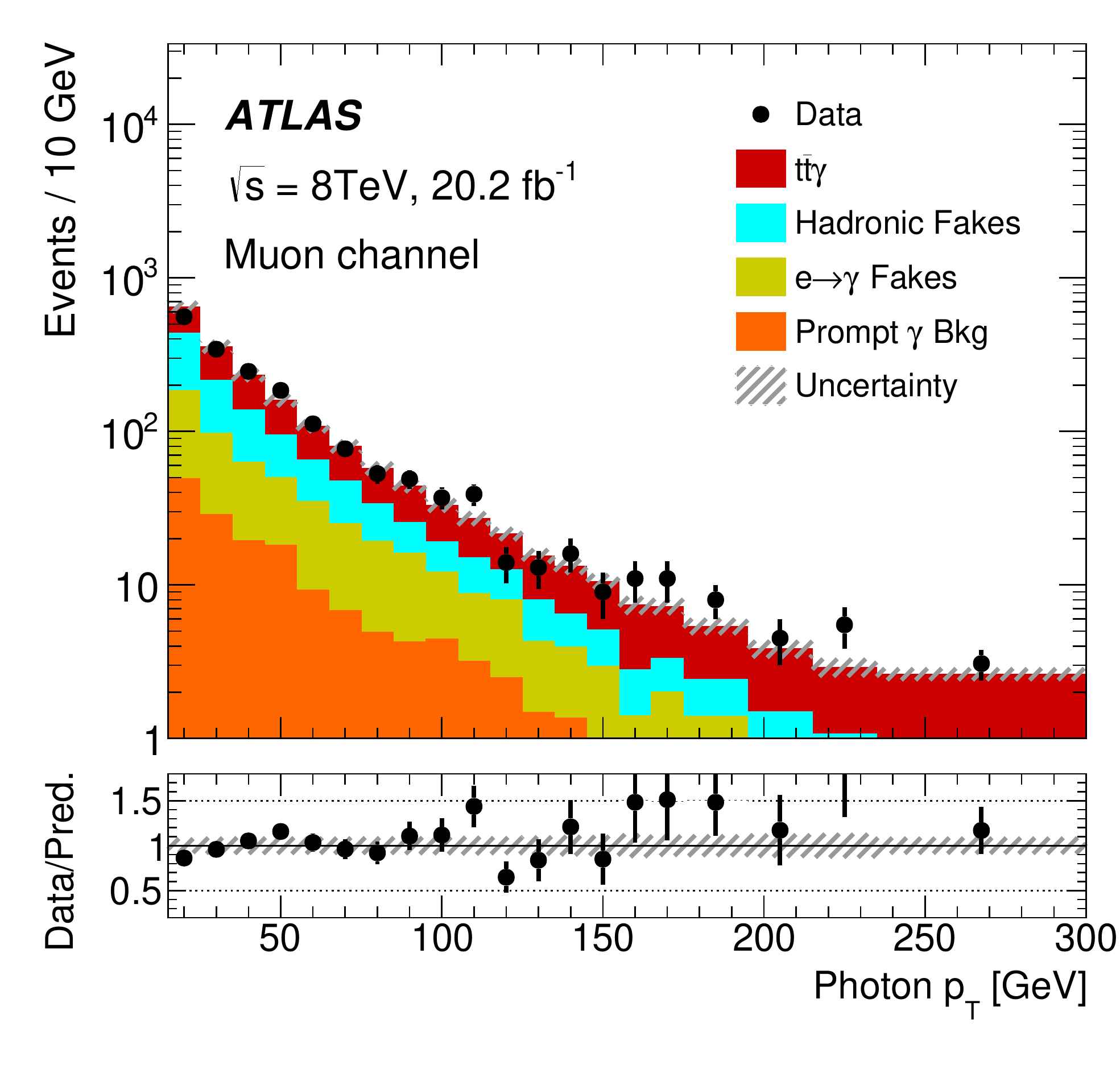}\label{fig:sel_mu_ph_pt}}
}
{
  \subfloat[]{\includegraphics[width=0.49\textwidth]{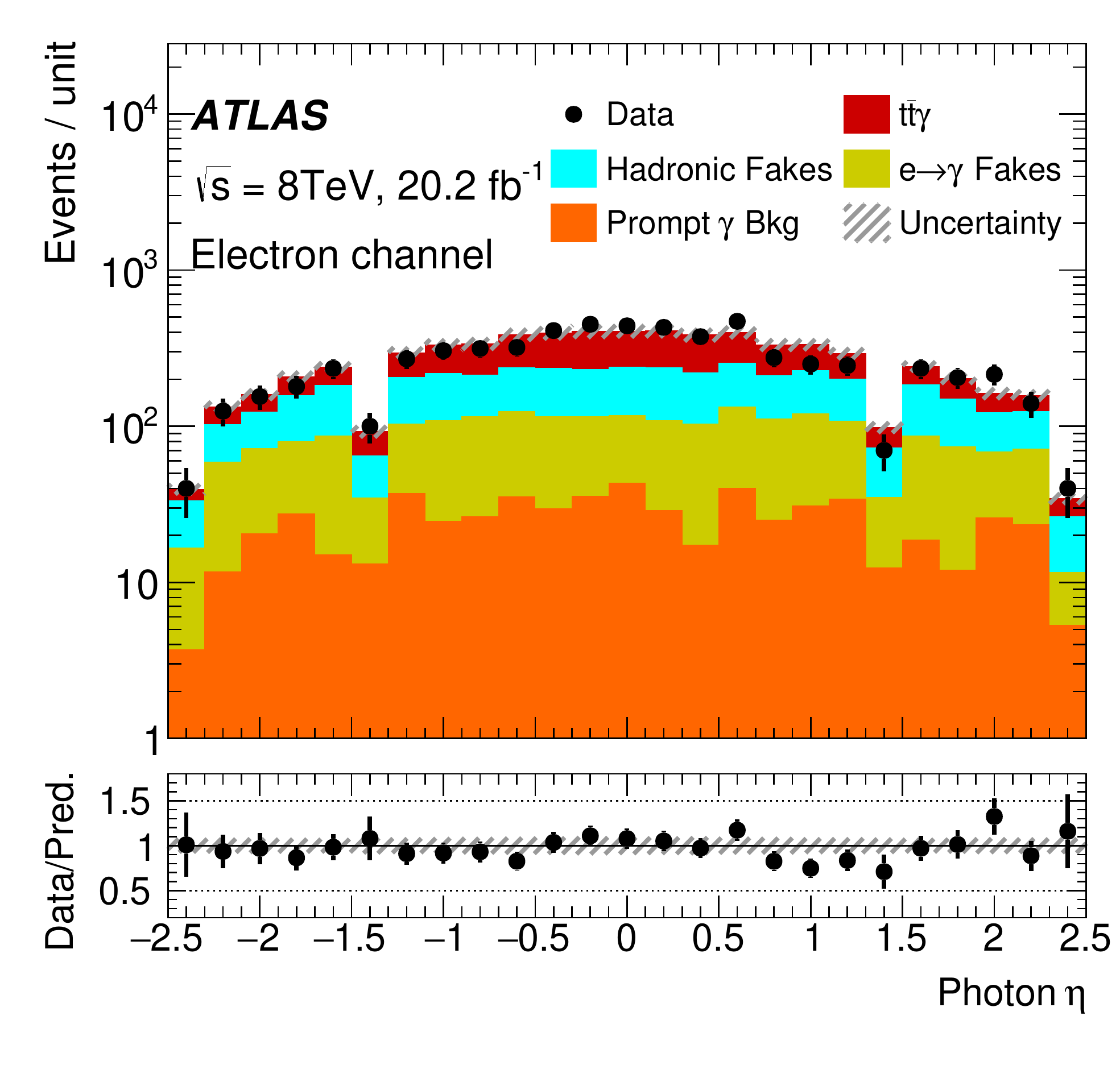}\label{fig:sel_el_ph_eta}}
  \subfloat[]{\includegraphics[width=0.49\textwidth]{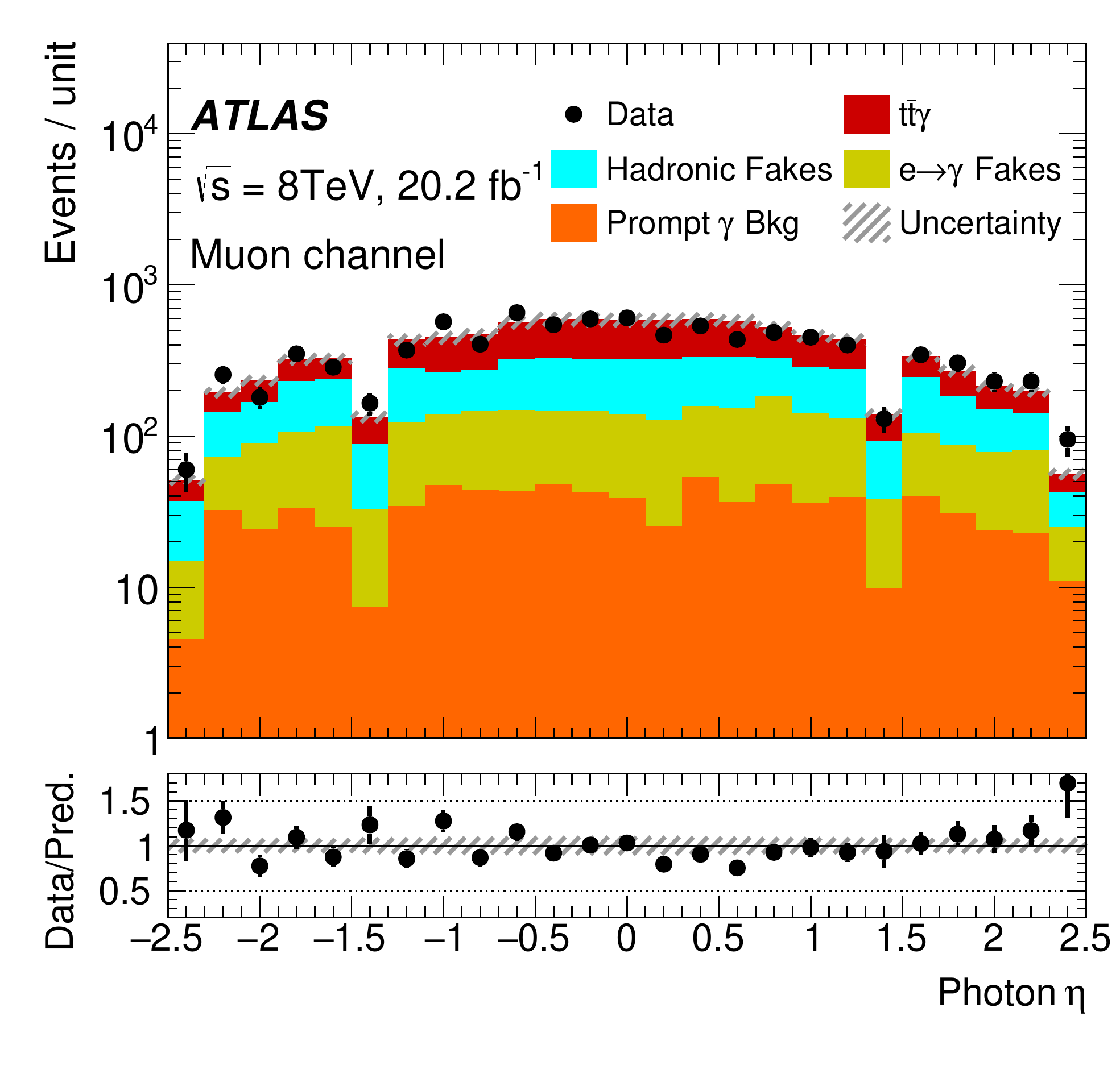}\label{fig:sel_mu_ph_eta}}
}
\caption[]{Comparison of data and the expected distributions in the events
passing the \ttg selection for the \subref{fig:sel_el_ph_pt},
\subref{fig:sel_el_ph_eta} single-electron channel and
\subref{fig:sel_mu_ph_pt}, \subref{fig:sel_mu_ph_eta} single-muon channel. The
expected signal and background distributions include the prediction for signal
(\ttg), normalised to the next-to-leading-order calculation~\cite{melnikov}, as
well as for other processes with prompt photons, such as $W\gamma$+jets and
$Z\gamma$+jets production from simulation (prompt $\gamma$ bkg), where the
$W\gamma$+jets contribution is scaled with a data-driven method.  The
contributions of events in which an electron is misidentified as a photon
($e\rightarrow\gamma$ fakes) and of events in which hadrons are misidentified
as photons (hadronic fakes) are estimated from data.  The shaded band
corresponds to the total uncertainty of the expected signal and backgrounds.
Panels \subref{fig:sel_el_ph_pt} and \subref{fig:sel_mu_ph_pt} show the photon
\pt distributions, where the overflow is included in the last bin, while
\subref{fig:sel_el_ph_eta} and \subref{fig:sel_mu_ph_eta} show the $\eta$
distributions of the photons.}
\label{fig:sel_slept} 
\end{figure}

The selection yields a total of 1256 and 1816 candidate events in the electron
and muon channels, respectively.  From simulation studies, $ 440 \pm 90$ and
$720 \pm 140 $ signal events are predicted in the electron and the muon
channels, respectively, including all systematic uncertainties discussed in
section~\ref{s:systematics}.  The photon \pt and \abseta distributions after
the selection are shown in figure~\ref{fig:sel_slept}.  The data distributions
are compared to the predictions from simulation and data-driven estimates for
both the signal and backgrounds, as explained in section~\ref{s:backgrounds}.

For the measurement of the differential cross sections, the phase space in the
variables \pT and $\eta$ is divided into five bins each, chosen to be
sufficiently large compared to the resolution in the respective variable and to
contain a similar number of predicted signal events. The bins in photon \pT are
$15$--\SI{ 25}{\GeV}, $25$--\SI{40}{\GeV}, $40$--\SI{60}{\GeV},
$60$--\SI{100}{\GeV}, and $100$--\SI{300}{\GeV}, and for the photon \abseta
they are $0$--$0.25$, $0.25$--$0.55$, $0.55$--$0.90$, $0.90$--$1.37$, and
$1.37$--$2.37$.  

\subsection{Definition of the fiducial phase space}
\label{subs:fiducial}

The fiducial region for this analysis is defined for Monte Carlo events at
particle level (before detector simulation) using the following particle
definitions and event selection, which are designed to mimic those at the
reconstruction level (after detector simulation).  The objects are constructed
from stable particles in the event record of the generator with a lifetime
larger than \SI{30}{ps}.

\begin{description}
\item [Leptons.] 
      Electrons and muons with $\pt > \SI{10}{\GeV}$ and $\abseta < 2.7$ are
      combined with all photons that do not originate from hadron decays and
      are within a cone of $\dR = 0.1$ and their four-momenta are added
      together.  These modified leptons are then required to have $\pt >
      \SI{25}{\GeV}$, $\abseta < 2.5$ and not originate from a hadron decay.
\item [Jets.]
      Jets are clustered with the anti-$k_t$ algorithm with a radius parameter
      of $R = 0.4$. Neutrinos and muons are not considered in the clustering.
      Jets are required to have $\pt > \SI{25}{\GeV}$ and $\abseta < 2.5$.
\item [$b$-jets.]
      Jets are tagged as $b$-jets if they contain a $b$-hadron with $\pT >
      \SI{5}{\GeV}$ within $\dR = 0.3$ of the jet axis.
\item [Photons.]
      Photons are required to not originate from a hadron decay and to satisfy
      $\ET > \SI{15}{\GeV}$ and $\abseta < 2.37$.
\item [Overlap removal.]
      The applied overlap removal procedure is the same as for the
      reconstructed objects, as described in section~\ref{s:objects}.
\end{description}

The following event selection based on the object definitions listed above is
applied to define the fiducial phase space: exactly one electron (muon) from a
$W$ boson decay is required in the electron (muon) channel; at least four jets
have to be selected, among which at least one must be a $b$-jet; and exactly
one photon is required. Additionally, the event is discarded if the photon has
$\Delta R(\text{jet} ,\gamma) < 0.5$ with any jet or $\Delta R(\ell,\gamma) <
0.7$ where $\ell$ is the electron or muon.

In order to obtain a common fiducial region for the electron channel and the
muon channel, the requirements imposed at the reconstruction level on \MET,
\mtw, and $m_{e\gamma}$ are not included in the fiducial region definition. 

\section{Templates}
\label{s:templates}

After the event selection there are three classes of events, 1) those with
prompt photons, 2) those with photons from hadron decays or hadrons
misidentified as photons, called hadronic fakes, and 3) those from electrons
misidentified as photons.  The prompt-photon category includes both the signal
events and other background processes, such as $W$ or $Z$ boson production with
prompt photons.  The extraction of the total and differential cross sections is
based on a likelihood fit using three templates, one for the prompt-photon
events, one for the hadronic-fake events and one for electrons misidentified as
photons.  The normalisations of the first two templates are free parameters in
the likelihood fit, while for the third template the normalisation is fixed to
the data-driven estimate of the number of events with an electron misidentified
as a photon, as described in section~\ref{subsec:egammaFakes}.  The variable
used for the templates is \ptcone, the sum of the transverse momenta of all
tracks within a cone with an opening angle around the photon of \SI{0.2}{rad}.
This variable yields the best discrimination between signal and background and
has almost no dependence on the amount of pile-up~\cite{TOPQ-2012-07}.  A
detailed description of the likelihood fit is given in section~\ref{s:fit},
while the background determination is discussed in section~\ref{s:backgrounds}.

\subsection{Prompt-photon template}
\label{s:sigtemplate}

\begin{figure}[htbp]
\centering
\includegraphics[width=0.8\textwidth]{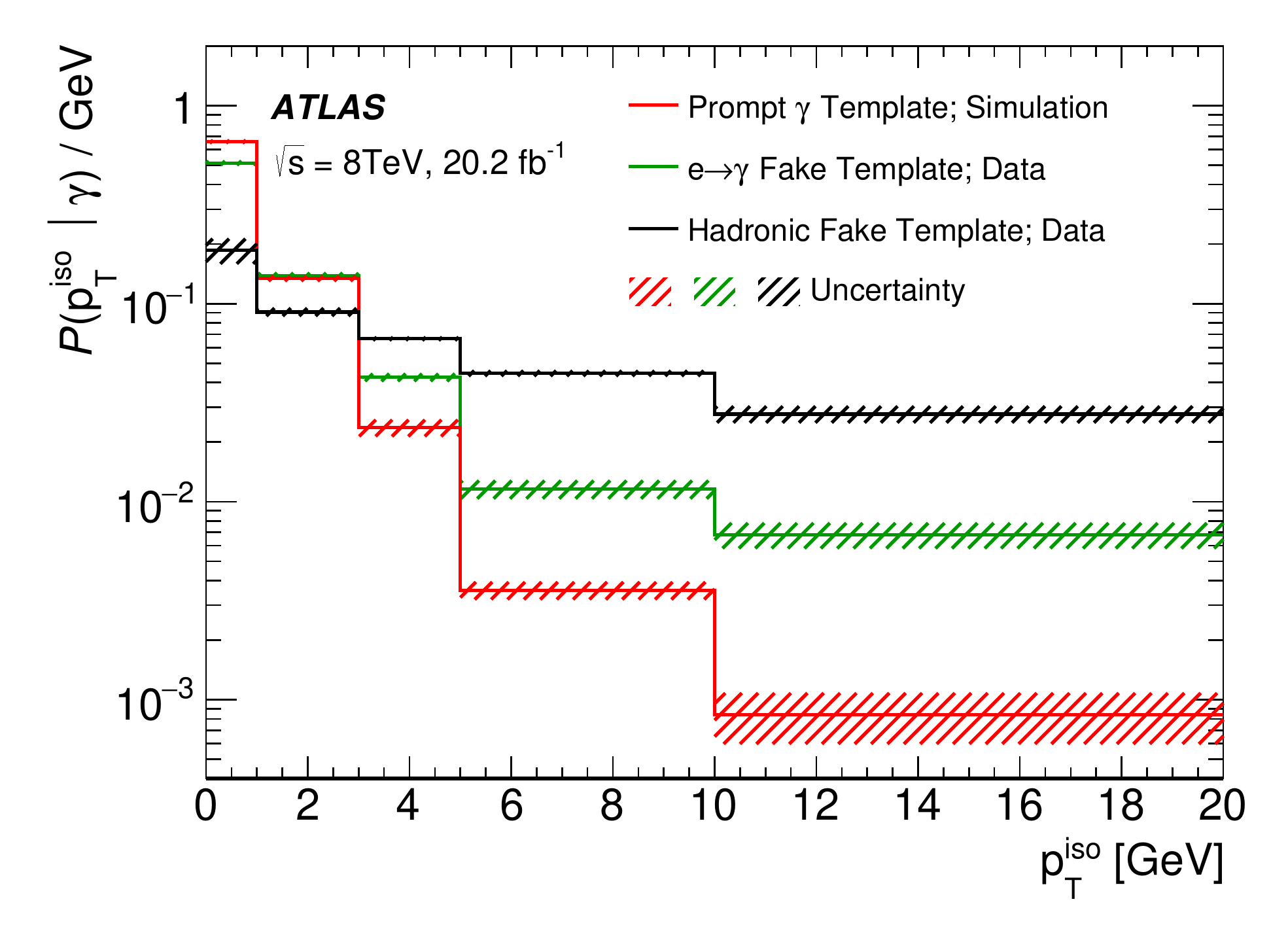}
\caption{The \ptcone templates for the inclusive cross-section measurement for
prompt photons, hadronic fakes and electron fakes.  The template for the 
signal photon is taken from simulation, while the other two templates are derived
from data.  The distributions are normalised to unity and the last bins contain
the overflows.  The shaded bands show the total uncertainty in each template.} 
\label{fig:Template}
\end{figure}

The prompt-photon template is extracted using the photons from the \ttg signal
simulation after the event selection described in section~\ref{s:selection}.
In addition, only reconstructed photons that are geometrically matched to a
particle-level photon within $\dR$ of 0.1 are selected. Approximately $95\%$ of
the photons in the selected signal events fulfil this condition.  The
prompt-photon template is shown in figure~\ref{fig:Template}.  Templates
extracted from $W \gamma + $jets and $Z \gamma + $jets simulation are
consistent with this template within statistical uncertainties.  The simulated
sample is large enough to ensure a small statistical uncertainty of the
template for the total cross-section measurement, with a maximum uncertainty of
$4\%$ in the last bin.  For the differential measurements, the template is
extracted for each bin of the transverse momentum and pseudorapidity
distributions of the photon. In this case, the uncertainty in the last bin of
the distribution can be as large as $13\%$.  Both the statistical and
systematic uncertainties, discussed in section~\ref{s:systematics}, are taken
into account when performing the likelihood fit described in
section~\ref{s:fit}. 

\subsection{Hadronic-fake template}
\label{s:hadronfake}

The hadronic-fake template is extracted from a control region in data with at
least four jets.  Events must have at least one photon candidate that fails to
satisfy at least one of the four photon identification criteria constructed
using shower-shape variables from the first layer of the EMC.  The strip layer
is finely segmented in $\eta$ for suppressing fake photons, which typically
have a broader shower profile.  While these variables have strong
discriminating power between signal and fake photons, they have negligible
correlation with the photon isolation~\cite{PERF-2013-04}.  The control region
is defined with the same jet multiplicity requirement as for the signal region
because the shape of the template depends on the jet multiplicity.  In
addition, to prevent electrons that are misidentified as photons from entering
the control sample, hadronic-fake photon candidates within $\dR = 0.1$ of an
electron are rejected.

The shape of the template depends on both the \pt and $\eta$ of the hadronic
fakes.  The hadronic fakes are less isolated at higher \pt values, as they are
more likely to arise from energetic jets which have larger \ptcone values.  The
dependence on $\eta$ is due to the varying amount of material in front of the
calorimeter.  Because of the different \pt and $\eta$ distributions of the
hadronic fakes in the control region and in the signal region, the
hadronic-fake template for the total cross-section measurement is calculated as
a weighted sum of templates  determined in bins of the hadronic fakes: five for
\pT (here, the same bins as for the photon \pT in the differential
cross-section measurement are chosen) and two for \abseta ($\abseta < 1.80$ and
$1.80 < \abseta < 2.37$).  The weights $w$ are the fraction of the
hadronic-fake events in a given \pt or $\eta$ range, estimated from another
control region in data, closer to the signal region although with fewer events,
in order to correct the \pt and $\eta$ distributions of the hadronic fakes.
This region is defined exactly in the same way as the signal region, described
in section~\ref{s:selection}, but replacing the nominal photon selection by one
where the photon fails to satisfy at least one of the tight identification
criteria.  The difference in the event topologies between the two control
regions was found to have a negligible effect on the final template shape.
Since no correlation between \pt and $\eta$ of hadronic fakes is observed in
this control region, the dependence of the template on \pt and $\eta$ can be
treated separately. 

The final weighted hadronic-fake template for the inclusive measurement, shown
in figure~\ref{fig:Template}, is calculated according to 

\begin{equation*}
  T^{\text{h-fake}}_{\text{weighted}} = \frac{1}{2}\left(\sum\limits_{i=1}^5
  w_{\pt,i}~~T^{\text{h-fake}}_{\pt,i} + \sum\limits_{j=1}^2
  w_{\eta,j}~~T^{\text{h-fake}}_{\eta,j}\right)\,,
\end{equation*}

where $T^{\text{h-fake}}_{\pT,i}$ is the hadronic-fake template of the \pt bin
$i$ and $w_{\pt,i}$ is its corresponding weight, while
$T^{\text{h-fake}}_{\eta,j}$ is the hadronic-fake template of the $\eta$ bin
$j$ with $w_{\eta,j}$ as its associated weight.  The index $i$ runs over the
five bins in \pT, while the index $j$ runs over the two bins in $\eta$.
Similarly, for each bin of the differential cross-section measurement in photon
\pT ($\eta$), a template is obtained by averaging over the hadronic-fake $\eta$
(\pT) dependence.

The main uncertainty in this template stems from a small contamination of the
selected events by prompt photons.

\subsection{Electron-fake template}
\label{s:electronfake}

Electrons misidentified as photons represent the second-largest background.
Electrons and photons have very similar shower shapes in the electromagnetic
calorimeter.  For this reason, an electron can be misidentified as a photon if
the associated track is poorly reconstructed, or in case of nearby jet activity
it can be misidentified  as a converted photon.  A \ptcone template is derived
from photons in a control sample containing electron-photon pairs with an
invariant mass compatible with the $Z$ boson mass, hereafter called
$Z\rightarrow e + \mathrm{fake}\: \gamma$ events. In these events the photon is
predominantly an electron misidentified as a photon.  The control sample
selection requires an electron and a photon with an opening angle larger than
$150^\circ$ and an invariant mass in the range $70$--\SI{110}{\GeV}.  The \pT
of the electron has to be larger than that of the photon.  In order to select
events with a topology close to the signal events, a requirement of \MET $>$
\SI{30}{\GeV} is applied.  Other backgrounds in this region are estimated from
a sideband fit to the $m_{e\gamma}$ distribution modelled by a Gaussian
function.  The corresponding template, after background subtraction, is shown
in figure~\ref{fig:Template}.

\section{Background estimation}
\label{s:backgrounds}

Several background processes can mimic the \ttg signature of the signal events.
In order of importance the main background contributions are fake photons from
misidentified hadrons, predominantly in \ttbar events
(section~\ref{subsec:hadFake}), and events with electrons misidentified as
photons (section~\ref{subsec:egammaFakes}).  Further small contributions stem
from processes with prompt photons (section~\ref{subsec:otherbkg}).  The sizes
of the background contributions are given in section~\ref{sec:result}.  

\subsection{Background from hadrons misidentified as photons}
\label{subsec:hadFake}

The main background contribution to the \ttg process comes from hadrons, or
photons from hadron decays, that are misidentified as prompt photons.  This
background is estimated from data using the template fit.  The main photon
background contribution is \ttbar production with one hadronic fake.  The derivation of
the template used for the hadrons misidentified as photons is described in
section~\ref{s:hadronfake}. 

\subsection{Background from electrons misidentified as photons}
\label{subsec:egammaFakes}

Events with electrons misidentified as photons are the second largest
background contribution.  The main contribution arises from \ttbar events with
both top quarks decaying semileptonically thereby producing two electrons or
one electron and one muon.  The second largest contribution is $Z$+jets
production where the $Z$ boson decays into an $e^+e^-$ pair and one of the
electrons is misidentified as a photon.  The contribution of events with
electrons misidentified as photons is estimated with a fully data-driven
method.

The probability for an electron to fake a photon (fake rate) is calculated
using two control regions, one enriched in \Zee and one enriched in events
reconstructed as $Z\rightarrow e + \mathrm{fake}\: \gamma$ events.  The event
selections for the \Zee and $Z\rightarrow e + \mathrm{fake}\: \gamma$ samples
require two back-to-back objects, either two electrons or an electron and a
photon. The definitions are the same as for the signal selection except for the
second electron in the \Zee selection, where the \pT threshold is lowered from
\SI{25}{\GeV} to \SI{15}{\GeV} so that it is the same as for the photon.  In
the   $Z\rightarrow e + \mathrm{fake}\: \gamma$ selection the electron is
required to have a larger \pT than the photon.  In both selections the object
with the larger \pT is the tag while the other one is the probe, i.e. the tag
is always an electron.

The numbers of \Zee and $Z\rightarrow e + \mathrm{fake}\: \gamma$ events are
determined from a fit of the invariant mass distribution of the two objects
using a sum of a Crystal-Ball~\cite{CrystalBall} and a Gaussian function.  The
fit is performed in the region where the invariant mass lies between
\SI{70}{\GeV} and \SI{110}{\GeV}.  The fake rate is calculated from the ratio
of these two numbers, as a function of the transverse energy and the
pseudorapidity of the photon.  A correction accounting for the different
reconstruction and identification efficiencies for electrons and photons is
applied. 

The contribution of electrons reconstructed as photons in the signal region is
then estimated by applying the fake rate to a modified signal region, where an
electron fulfilling the same kinematic selection as the photon is required.
This background contributes $317 \pm 7 \pm 41$ and $385 \pm 6 \pm 42$ events in
the electron and muon channels, respectively, where the first error is
statistical and the second is systematic.  The systematic uncertainty is
estimated by varying the range and functions used in the fit of the two
invariant mass distributions. 

\subsection{Background events with a prompt photon}
\label{subsec:otherbkg}

There are several background processes to \ttbar production that can become a
background to the \ttg sample, if an additional prompt photon is present.

Multijet production with an associated prompt photon is a source of background
when one of the jets in the event is misidentified as a lepton, referred to as
a fake lepton.  In order to estimate this background, a control sample is
created that uses the same event selection as for the signal except that the
lepton identification criteria are loosened.  To account for the sample
composition in fake and real leptons, each event is assigned a weight computed
from the matrix method~\cite{TOPQ-2010-01} in order to obtain a distribution
corresponding to fake leptons.  To calculate the contribution of events with a
prompt photon in this sample, a likelihood fit is performed on the \ptcone
distribution of these weighted events, using the prompt-photon and hadron-fake
templates described in section~\ref{s:sigtemplate} and
section~\ref{s:hadronfake}, respectively. This results in estimated background
contributions of $7.5 \pm 3.6$ events in the electron channel and $8.3 \pm 5.2$
events in the muon channel.  The uncertainties are purely statistical. 

Additional backgrounds from $W \gamma + $jets, $Z \gamma + $jets, single top
quark and diboson production are estimated using Monte Carlo simulations.  For
the $W \gamma + $jets events, the cross sections from simulation do not agree
with the data, as the fraction of heavy-flavour jets observed in data is
smaller than the prediction obtained using the \SHERPA generator, and thus a
scale factor is applied to the prediction of the simulation.  This scale factor
is derived from a control region in data and is $0.69 \pm 0.16$ for the
electron channel and $0.76 \pm 0.14$ for the muon channel.  The control region
is defined exactly as for the signal region, except for the required number of
jets (between one and three), the number of $b$-tagged jets (exactly one), and
for an additional requirement on the invariant mass of the lepton and the
photon, which has to be less than \SI{40}{\GeV}.  The uncertainties include the
statistical and systematic uncertainty induced by the subtraction of the non-$W
\gamma + $jets events from the control region. A comparison between \ALPGEN and
the nominal generation with \SHERPA gives an additional $20\%$ uncertainty.
For the $Z\gamma + $jets, single-top and diboson production with an additional
photon, a $50\%$ theoretical uncertainty is assumed.  In addition, all
uncertainties discussed in section~\ref{s:systematics} are taken into account.

The expected yields for these backgrounds are summarised in
table~\ref{tab:promptBkg}.

\begin{table}[htbp]
\centering
\begin{tabular}{|lcc|}
\hline
Process             & $e$ + jets    & $\mu$ + jets \\
\hline
Multijet + $\gamma$  & $7.5 \pm 3.6$ & $8.3 \pm 5.2$ \\
$W\gamma +$jets         & $65 \pm 25$   & $97 \pm 25$ \\
$Z\gamma +$jets         & $35 \pm 19$   & $38 \pm 20$ \\
Single top + $\gamma$   & $13 \pm 7$    & $19 \pm 10$ \\
Diboson + $\gamma$      & $2.6 \pm 1.5$ & $2.5 \pm 1.4$ \\
\hline
\end{tabular}
\caption[]{Expected yields of background processes with a prompt photon.  The
uncertainties include all sources of systematic uncertainty described in
section~\ref{s:systematics}.}
\label{tab:promptBkg}
\end{table}

\section{Measurement of the fiducial cross section}

In this section, the theoretical predictions for the fiducial region \ttg cross
section and differential cross sections are described.  The fit strategy used
to extract these from the data is also presented.

\subsection{Theoretical prediction}
\label{subs:theory}

Production cross sections of a top-quark pair in association with a prompt
photon have been calculated at NLO in QCD~\cite{melnikov}, extending
calculations obtained assuming stable top quarks~\cite{duan1}.  While the
results presented in ref.~\cite{melnikov} were calculated at a centre-of-mass
energy of $\sqrt{s} = \SI{14}{\TeV}$, a dedicated calculation at $\sqrt{s} =
\SI{8}{\TeV}$ has been performed in the single-lepton channel both at LO and
NLO using the same techniques as in ref.~\cite{melnikov}.  The \pdf{CTEQ6L1}
(\pdf{CT10}) PDF set is used for the LO (NLO) calculation, while for both a
fine-structure constant $\alpha= 1/137$ and renormalisation and factorisation
scales $\muren = \mufac = \mtop$ are used.  Further computations were performed
by simultaneously varying the scales \muren and \mufac by a factor of $2$ or
$0.5$.  The LO calculation agrees with the cross section from \MADGRAPH within
$2\%$, for scale choices of both $\mtop$ and $2\mtop$.  The ratio of the NLO
and LO cross sections, known as the $K$-factor, is obtained from the NLO theory
calculation at a scale of $\mtop$ and the LO \MADGRAPH sample at a scale of
$2\mtop$, after applying the restricted phase space requirements
described in section~\ref{s:samples}.
The overall $K$-factor was calculated to
be $K = 1.90 \pm 0.25 \pm 0.12$, with the first uncertainty accounting for
scale variation and the second for PDF set variation.  For the differential
cross-section predictions, the $K$-factors for the corresponding bin in \pt or
$\eta$ are used. These $K$-factors are used to correct the prediction from
simulation with \MADGRAPH.

The $K$-factor derived above is then used to calculate the cross section at NLO
by using only events from the simulation with \MADGRAPH in the same fiducial
region as defined in section~\ref{subs:fiducial}.  The inclusive cross section
is calculated to be \SI{151 \pm 24}{fb}.  In addition, the cross section is
calculated in bins of \pt and $\eta$ with the same fiducial requirements as for
the cross-section measurement.  

\subsection{Fit strategy} 
\label{s:fit}

The cross section for \ttg production is extracted from the observed binned
\ptcone distribution employing a likelihood fit.  There are two free parameters
in the fit, i.e. the total number of signal events and the total number of
hadronic-fake events.  The \ptcone templates used in the fit are shown in
figure~\ref{fig:Template}.  The prompt-photon template is used for the signal
and all backgrounds with a prompt photon.  For the background from electrons
misidentified as photons, the electron-fake template is utilised and for
hadronic fakes the hadronic-fake template is utilised.  All backgrounds, except
for the hadronic fakes, are described by a Poisson probability distribution
with the mean values determined in section~\ref{s:backgrounds}.

The likelihood function $\mathcal{L}$ is defined as the product

\begin{equation}
  \mathcal{L} = \prod_{i, j} P (N_{i,j}|N_{i,j}^{s} +
  \raisebox{1pt}{\ensuremath{\sum\limits_{b}}} N_{i,j}^{b}) \cdot \prod_{t} G ( 0
  |\theta_t,1).
\label{eqn:likelihood}
\end{equation}

The Poisson function $P(N_{i,j}|N_{i,j}^{s} + \sum_{b} N_{i,j}^{b})$ models the
event yield in bin $j$ of the \ptcone distribution of bin $i$ of the \pT or
$\eta$ distribution, where $N_{i,j}$ is the observed number of events and
$N_{i,j}^s$ and $N_{i,j}^{b}$ are the expected numbers of signal and background
events.  There is only one bin for the inclusive measurement, while there are
five bins for each of the differential cross-section measurements,
corresponding to the bins in photon \pt or $\eta$.  The Gaussian function $G( 0
|\theta_t,1)$ of unit width models the systematic uncertainty $t$, where
$\theta_t$ is the parameterisation of this uncertainty.  The sources of
systematic uncertainty are discussed in section~\ref{s:systematics}.

The inclusive and differential fiducial cross sections are related to the
number of signal events by

\begin{equation*}
  L \cdot \sigma_i \cdot C_i \cdot f_{i,j} = N_{i,j}^s,
\end{equation*}

where $L$ is the integrated luminosity of the data sample, $\sigma_i$ is the
fiducial cross section to be determined, $f_{i,j}$ is the fraction of events
falling into bin $j$ of \ptcone of bin $i$, calculated from the signal
template, and $C_i$ is the ratio of the number of reconstructed events to the
number of generated events in the fiducial region in bin $i$.  The term $C_i$
can also be expressed as the ratio of the number of reconstructed events to the
number of reconstructed events falling into the fiducial region in bin $i$,
multiplied by the signal efficiency.  The ratio $C_i$ therefore corrects for
the event selection efficiency and the migration of events between the fiducial
and the non-fiducial region, including leptonic $\tau$ decays,
or between different bins $i$, in the case of the
differential cross-section measurement.  The values of $C_i$ range from $0.21$
to $0.38$ ($0.38$ to $0.62$) for the electron (muon) channel, depending on the
bin in \pT or $\eta$.  For the differential cross-section measurements, this
means that the cross section $\sigma_i$ is computed in each bin $i$ using a
bin-by-bin unfolding.

To consider the effect of a systematic uncertainty $t$ on the value of a
parameter $p$ in the fit, $p$ is multiplied by a response function, which is a
function of the nuisance parameter $\theta_t$ used to parameterise this
uncertainty.  The parameters are the variables used to calculate $ N_{i,j}^s$
or $N_{i,j}^{b}$,

\begin{equation*}
  p(\theta_t) = p \cdot (1 + \sigma_p)^{\theta_t},
\end{equation*}

where $\sigma_p$ is the uncertainty of parameter $p$ due to the systematic
uncertainty $t$.  There could be multiple response functions multiplying $p$,
if it is affected by more than one systematic uncertainty.

The fiducial cross sections are extracted by maximising the likelihood function
(eq.~(\ref{eqn:likelihood})).  To obtain the confidence interval for the fitted
cross sections, a profile likelihood ratio $\lambda_s$ is built according to

\begin{equation*}
  \lambda_s (\sigma) = \frac{\mathcal{L}(\sigma, \hat{\hat{\MBF\theta}})}
  {\mathcal{L}(\hat{\sigma}, \hat{\MBF\theta})},
\end{equation*}

where the nuisance parameters are denoted by $\MBF\theta$.  The quantities with
single hats are their unconditional maximum-likelihood estimate, while the
quantities with double hats are their conditional maximum-likelihood estimate
when $\sigma$ is fixed. The profile likelihood ratio is evaluated within the
RooFit/RooStats framework~\cite{RooFit,RooStats} and used to determine the
upper and lower limits on the cross section at the $68\%$ confidence level.  In
the fit, events from the electron and muon channels are merged, having a common
parameter of interest, i.e. the fiducial cross section.

For the differential cross-section measurement, the fiducial region definition
includes all the requirements for the total cross-section measurement.  In
addition, the true \pT or $\eta$ of the photon has to be in the bin considered.
The systematic uncertainties are treated as correlated between the bins.  The
size of the bins is chosen to keep the migration between bins lower than the
expected statistical uncertainty of the cross-section measurement in that bin,
in order to have only a small contribution from the bin-by-bin unfolding to the
total uncertainty. The migration between bins is smaller than $7\%$ for all
bins.
 
\section{Systematic uncertainties}
\label{s:systematics}

There are three categories of systematic uncertainty affecting the results: the
modelling uncertainties, the experimental uncertainties, and the uncertainties
related to the template shapes.  When varying the parameters to estimate
uncertainties, only small differences between the positive and negative
uncertainties in the cross section are observed. The uncertainty is therefore
symmetrised by taking the larger of the two values.  The uncertainties for the
inclusive and differential measurements are derived in the same way.  The
uncertainties for the inclusive measurement are discussed in detail below  and
summarised in table~\ref{tab:sys}.  The uncertainties for the differential
measurements are included in table~\ref{tab:numresult}.

\subsection{Modelling uncertainties}
\label{s:modelling}

The renormalisation and factorisation scales for the simulation of the signal
process are varied simultaneously from their nominal value of $\muren = \mufac
= 2\mtop$ by a factor of $1/2$ or $2$, resulting in an uncertainty of $0.6\%$
in the measured cross section.  \PYTHIA~6, which is used for the nominal
signal sample, is replaced by \HERWIG~6.520~\cite{herwig} and
\textsc{Jimmy}~4.31~\cite{jimmy} to estimate the uncertainty due to the
modelling of the parton shower, underlying event, and hadronisation.  This
results in an uncertainty of $0.6\%$ in the cross section.  Initial- and
final-state radiation are studied by using \PYTHIA~6 tunes with high
(\tune{Perugia2011C radHi}) and low (\tune{Perugia2011C radLo}) QCD radiation
activity for the signal sample.  
This results in an uncertainty in the inclusive cross section of
$2.2\%$.

Other signal modelling uncertainties were studied, including the choice of
matrix-element event generator~\cite{TOPQ-2012-07}, the PDF, and the effect of
colour reconnection, underlying event, and QED uncertainties, and are found to
be negligible.  The uncertainties in the $Z\gamma + $jets, single-top$+ \gamma$
and diboson$+ \gamma$ backgrounds are estimated using the 48\% uncertainty in
the normalisation of the samples in the four-jet bin from the Berends--Giele
scaling~\cite{Berends}.  For the $Z\gamma +$jets events, this is the largest
uncertainty, resulting in an uncertainty of $2.8\%$ in the inclusive
cross-section measurement. For the single-top$+ \gamma$ events the uncertainty
is $1.2\%$, while for the diboson$+ \gamma$ events the uncertainty is
negligible.

For $W\gamma +$jets events, the uncertainty in the scale factor used to
normalise the sample (section~\ref{subsec:otherbkg}), as well as the difference
between the predictions from \SHERPA and \ALPGEN, is taken into account,
resulting in a total uncertainty of $4.0\%$.

The fake rate used to estimate the background from electrons misidentified as
photons shows a dependence on the choice of range and function used in the fit
of the $ee$ and $e\gamma$ invariant mass distributions, resulting in a total
systematic uncertainty of $6.1\%$ in the measured inclusive cross section.

\subsection{Experimental uncertainties}
\label{s:experimental}

Experimental uncertainties common to signal and background processes come
mainly from the uncertainties associated with the event reconstruction,
identification, and trigger efficiencies, momentum and energy scales, and
momentum and energy resolutions of the jets, the photon, the lepton, and \MET.
In addition, the uncertainties associated with the jet flavour tagging, the
integrated luminosity, and the pile-up simulation are considered.

The largest uncertainty associated with jets arises from the jet energy scale
(JES)~\cite{PERF-2012-01}, which is split into several independent categories.
To determine the JES uncertainty, each source is varied independently, and the
results added in quadrature to obtain an uncertainty in the inclusive cross
section of $4.9\%$.  The leading sources for the JES uncertainty are due to
uncertainties in the modelling, the amount of pile-up, and the jet flavour
composition.  The jet energy resolution is evaluated in a similar
way~\cite{PERF-2011-04} and results in a total uncertainty of $0.5\%$.

The photon identification efficiency~\cite{PERF-2013-04} is measured with
samples of photons from the radiative decays of the $Z$ boson, and electrons
and positrons from $Z$ boson decays, exploiting the similarity between electron
and photon electromagnetic showers.  Scale factors are used to correct for
detector mismodelling.  Uncertainties in these scale factors range between
$1.5\%$ and $2.5\%$ ($2\%$ and $3\%$) for unconverted (converted) photons in
the region of $\ET < \SI{40}{\GeV}$, and $0.5\%$ to $1\%$ for higher transverse
energies.  These scale factors are varied to study their impact on the
analysis, resulting in an uncertainty in the inclusive cross section of
$1.2\%$.  The photon energy scale uncertainty~\cite{PERF-2013-05} contributes
another $0.7\%$ uncertainty to the measured inclusive cross section.

For leptons, the trigger and identification efficiencies, and the energy scale
and resolution have been investigated~\cite{Aaboud:2016vfy, PERF-2011-04,
PERF-2014-05}.  Correction factors that are applied to the simulation to better
match the data are varied within their uncertainty, resulting in an uncertainty
of $1.1\%$ in the total cross-section measurement.

The uncertainty in $b$-tagging~\cite{PERF-2012-04} is accounted for by varying
the calibration scale factors for $b$-jets, $c$-jets, and light-flavour jets by
their corresponding systematic uncertainties independently yielding a total
systematic uncertainty of $0.3\%$.  The contribution due to the \MET
uncertainty is negligible.

The uncertainty in the integrated luminosity is $1.9\%$, as determined from a
calibration of the luminosity scale derived from beam-separation
scans~\cite{DAPR-2013-01}.  As the luminosity affects both the signal and
background normalisation the resulting uncertainty is $2.1\%$.  

\subsection{Template-related uncertainties} 

The dominant uncertainty in the hadron-fake template comes from the
contamination by prompt photons.  In order to take this contamination into
account, a modified template is constructed, where the photon candidates are
required not only to fail one of the photon identification requirements based
on shower shapes, but all four of them (section \ref{s:hadronfake}).  The
likelihood fit is then repeated using the modified template resulting in a
systematic uncertainty in the inclusive cross section of $6.3\%$.  

The prompt-photon template is affected by the modelling and experimental
systematic uncertainties as described in sections~\ref{s:modelling}
and~\ref{s:experimental}.

The largest uncertainty due to the template describing electrons misidentified
as photons comes from a variation on the \MET requirement in the event
selection. Other uncertainties are due to the variation of the mass range in
the event selection and the \pT ordering chosen in the determination of the
electron fake rates. The combination of these effects, together with the fake
rate uncertainty introduced above, yields a systematics uncertainty of $6.3\%$
on the cross section measurement.

\begin{table}[htbp]
\centering
\begin{tabular}{|lc|}
\hline
Source & Relative uncertainty [\%] \\ 
\hline
Hadron-fake template & 6.3 \\
$e\to\gamma$ fake & 6.3 \\
Jet energy scale & 4.9 \\
$W\gamma+$jets & 4.0 \\
$Z\gamma+$jets & 2.8 \\
Initial- and final-state radiation & 2.2 \\
Luminosity & 2.1 \\
Photon & 1.4 \\
Single top$+\gamma$ & 1.2 \\
Muon & 1.2 \\
Electron & 1.0 \\
Scale uncertainty & 0.6 \\
Parton shower & 0.6 \\ 
\hline
Statistical uncertainty & 5.1 \\ 
\hline
Total uncertainty & 13 \\ 
\hline
\end{tabular}
\caption[]{List of the most important systematic uncertainties for signal and 
background and their effects on the measurement of the inclusive cross section.  
In addition, the statistical and total uncertainty are given.  The total uncertainty 
is derived from the likelihood fit.}
\label{tab:sys}
\end{table}

\clearpage
\section{Results}
\label{sec:result}

A total of $3072$ candidate events are observed in data.  The result of the fit
in the fiducial region for the inclusive cross section is summarised in
table~\ref{tab:numresult}, together with the event yields of the backgrounds.
No significant pull of the fit input values and their uncertainties is observed
compared to the post-fit values.  Also, the fitted event yields for the
differential cross-section measurement are shown in each photon \pT and $\eta$
bin.

\begin{table}[htbp]
\centering
\scalebox{0.78}{
\centering
\begin{tabular}{l*{8}{r@{ $\pm$ }l}r}
\hline
Range & \multicolumn{2}{c}{\ttg} & \multicolumn{2}{c}{Hadronic} & 
\multicolumn{2}{c}{$e \to \gamma$} & \multicolumn{2}{c}{$W\gamma +$jets} &
\multicolumn{2}{c}{$Z\gamma +$jets} & \multicolumn{2}{c}{Single} & 
\multicolumn{2}{c}{Multijet$+ \gamma$} & \multicolumn{2}{c}{Diboson$+ \gamma$}
& Data \\[-0.7ex] 
\multicolumn{3}{c}{} & \multicolumn{2}{c}{fake} & \multicolumn{2}{c}{fake} &  
\multicolumn{4}{c}{} & \multicolumn{2}{c}{top$+ \gamma$}  \\
\hline
             Total       &1060 & 130 & 1020 & 90    & 710 & 90    & 160 & 40    & 73 & 32    & 32  & 15    & 16 & 6  & 5.1 & 2.4      & 3072\\
\hline
$15 \leq \pt < 25$ \GeV  & 280 & 40 & 360  & 40    & 240 & 35    & 47  & 13    & 23 & 10    & 7   & 4     & 4.4 & 2.3   & 1.3 & 0.7       & 966\\
$25 \leq \pt < 40$ \GeV  & 309 & 34 & 233  & 26    & 171 & 7     & 37  & 10    & 22 & 10    & 6.4 & 3.3   & 3.8 & 2.4    & 1.8 & 0.9      & 783\\
$40 \leq \pt < 60$ \GeV  & 220 & 40 & 205  & 21    & 111 & 30    & 28  & 8     & 13 & 6     & 10  & 5     & 1.6 & 1.9   & 0.5 & 0.3       & 589\\
$60 \leq \pt < 100$ \GeV & 160 & 40 & 116  & 16    & 100 & 40    & 24  & 7     & 10 & 5     & 8   & 4     & 3.4 & 2.1   & 1.0 & 0.6       & 420\\
$100 \leq \pt < 300$ \GeV& 150 & 25 & 71   & 10    & 50  & 20    & 23  & 7     & 4  & 2     & 0.9 & 0.7   & 0.8 & 1.0  & 0.3 & 0.2        & 298\\
\hline
      $ \abseta < 0.25$   & 246 & 34 & 121  & 21    & 93  & 24    & 18  & 6     & 9  & 4     & 4.0 & 2.2   & 5.2 & 1.8  & 1.0 & 0.6        & 497\\
$0.25 \leq \abseta < 0.55$& 260 & 40 & 130  & 20    & 116 & 29    & 29  & 8     & 11 & 6     & 3.7 & 2.1   & 0.0 & 0.4   & 1.5 & 0.8       & 552\\
$0.55 \leq \abseta < 0.90$& 180 & 40 & 198  & 27    & 150 & 40    & 31  & 9     & 16 & 7     & 2.2 & 1.3   & 4.0 & 1.8   & 0.4 & 0.2       & 578\\
$0.90 \leq \abseta < 1.37$& 200 & 40 & 233  & 33    & 169 & 50    & 35  & 10    & 17 & 8     & 9   & 5     & 5.7 & 2.1   & 1.0 & 0.5       & 663\\
$1.37 \leq \abseta < 2.37$& 150 & 40 & 344  & 33    & 200 & 12    & 48  & 13    & 19 & 9     & 13  & 6     & 5.4 & 2.5   & 1.4 & 0.7       & 782\\
\hline
\end{tabular}
}
\caption[]{Post-fit event yields for the signal and backgrounds for the
inclusive cross-section measurement and for the different bins of reconstructed
photon \pT and $\eta$ for the differential cross-section measurement.  The
uncertainties include the statistical and systematic uncertainties.}
\label{tab:numresult}
\end{table}

\begin{figure}[htbp]
\centering
{
\subfloat[]{\includegraphics[width=0.52\textwidth]{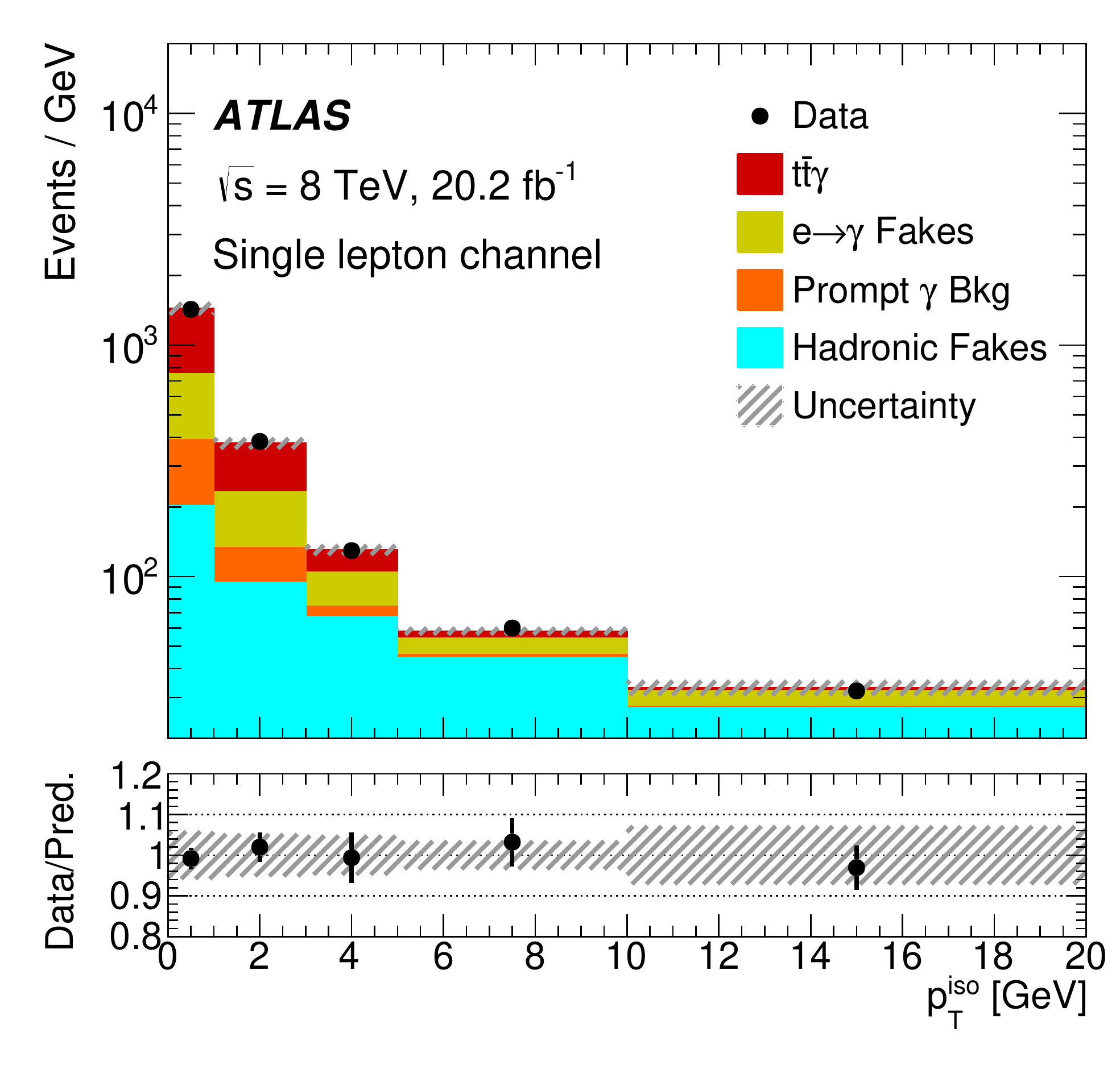}\label{fig:post_ptcone20_fidu}}
\subfloat[]{\includegraphics[width=0.47\textwidth]{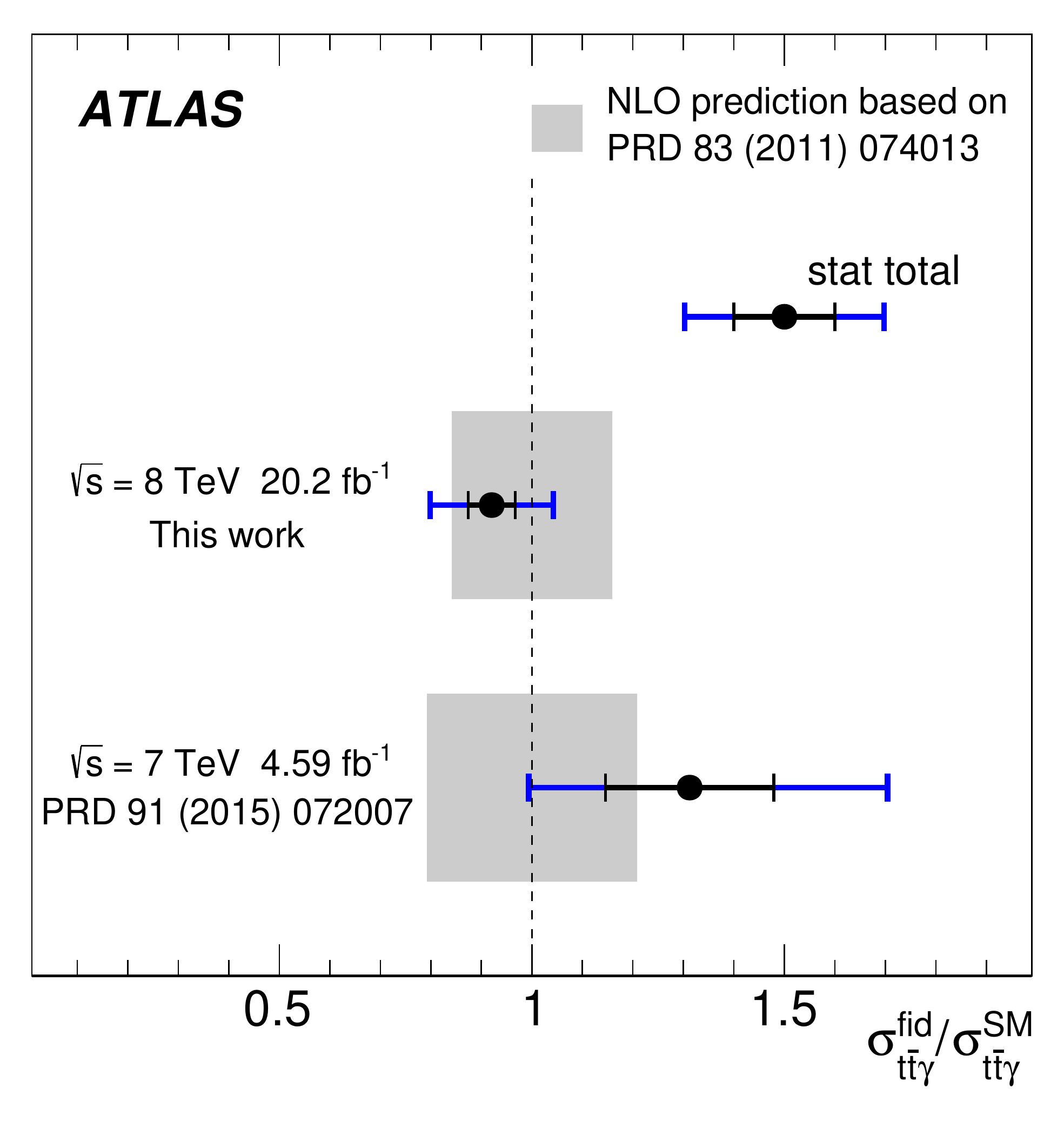}\label{fig:summary}}
}
\caption[]{\subref{fig:post_ptcone20_fidu} Post-fit track isolation
distribution for the inclusive cross-section measurement. The last bin includes
the overflow.  The uncertainty band includes all uncertainties.
\subref{fig:summary} Summary of fiducial measurements of \ttg production in
$pp$ collisions at \SI{7}{\TeV}~\cite{TOPQ-2012-07} and \SI{8}{\TeV},
normalised to the expected cross section, calculated at NLO
accuracy~\cite{melnikov}.} 
\end{figure}

The post-fit track isolation distribution for the inclusive measurement is
shown in figure~\ref{fig:post_ptcone20_fidu}.

Using data selected in the single-lepton channel, the result of the inclusive
measurement is: 

\begin{equation}
 \sffl = 139  \pm  7  \stat \pm  17 \syst~\text{fb} = 139 \pm  18~\text{fb},
 \nonumber
\end{equation}

which agrees with the NLO prediction of $151 \pm 24$ fb~\cite{melnikov}.  This
result is compared to the Standard Model calculation as well as to the
measurement performed at $\sqrt{s} = \SI{7}{\TeV}$~\cite{TOPQ-2012-07} in
figure~\ref{fig:summary}.  Good agreement with the NLO predictions is observed
in both cases.

\begin{figure}[htbp]
\centering
{
\subfloat[]{\includegraphics[width=0.49\textwidth]{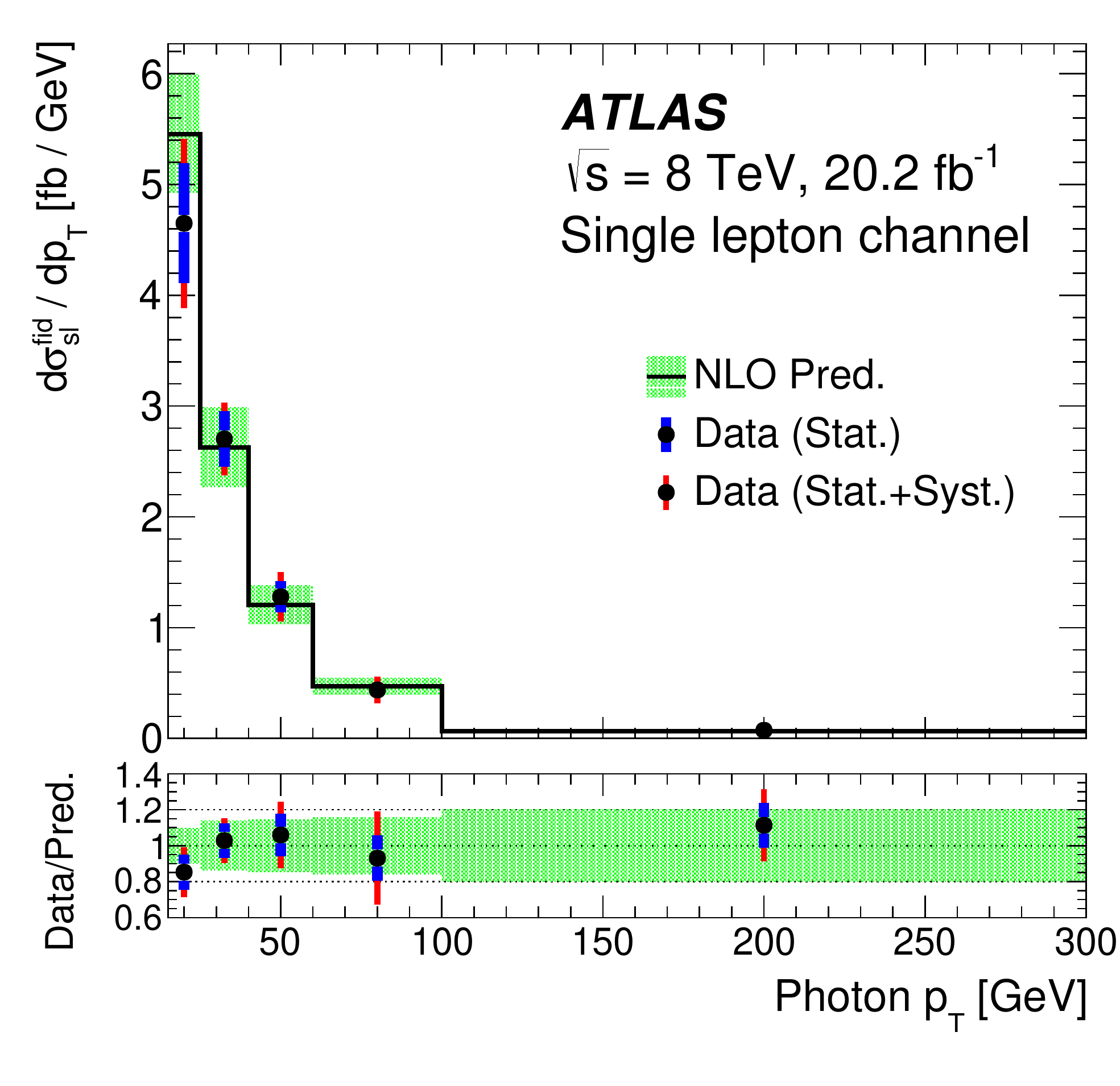}\label{fig:diff_xsec_a}}
\subfloat[]{\includegraphics[width=0.49\textwidth]{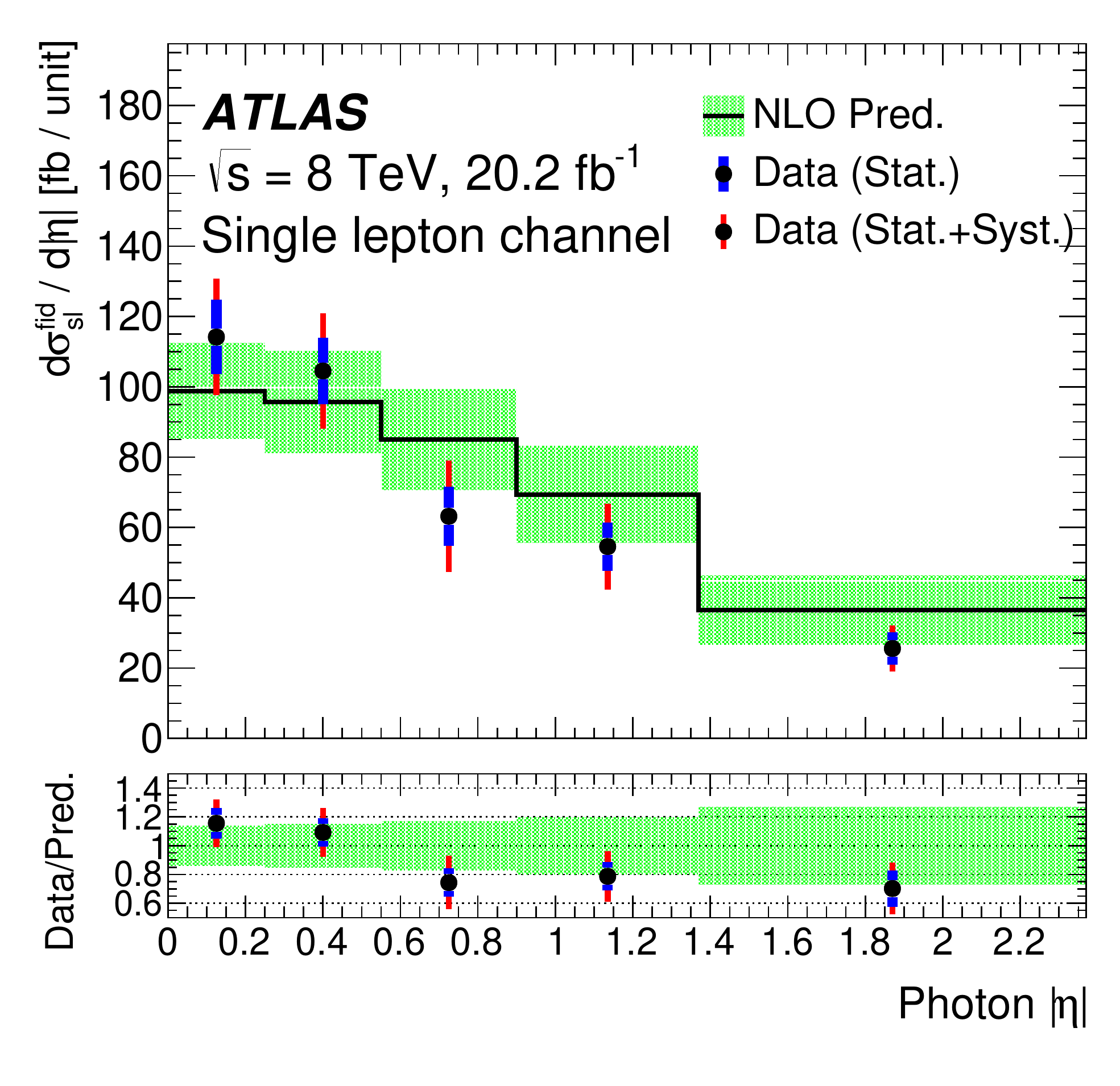}\label{fig:diff_xsec_b}}
}
\caption[]{Measured differential cross section in \subref{fig:diff_xsec_a}~\pT
and \subref{fig:diff_xsec_b}~\abseta and the corresponding theoretical
prediction.} \label{fig:diff_xsec}
\end{figure}

The measured \pT and $\eta$ differential cross sections are shown and compared
with their corresponding theoretical predictions in figure~\ref{fig:diff_xsec}.
Good agreement is observed between the measurement and predicted values.

\clearpage
\section{Conclusions}
\label{s:conclusion}

A measurement of the fiducial cross section of top-quark pair events in
association with a photon, using \SI{20.2}{\ifb} of
\SI{8}{\TeV} $pp$ collision data collected in 2012 by the ATLAS detector at the
LHC is presented.  The measurement is performed in the single-lepton channel
with either one isolated electron or muon, at least four jets where at least
one is $b$-tagged in the fiducial region $\pt(\gamma) > \SI{15}{\GeV}$,
$|\eta(\gamma)| < 2.37$ and $\pt(\ell) > \SI{25}{\GeV}$, $|\eta(\ell)| < 2.5$,
with angular separations $\Delta R(\ell,\gamma) > 0.7$ and $\Delta
R(\text{jet},\gamma) > 0.5$.  The result is based on the minimisation of a
profile likelihood ratio, using the photon track isolation as the
discriminating variable since it provides good rejection against the main
background of hadronic jets misidentified as photons.  The analysis uses
data-driven measurements for the determination of the largest background
contributions.  In particular, the track isolation template of the hadronic
fakes and the background contributions of electron fakes and QCD multijet
events are obtained from data.

The fiducial \ttg cross section is extracted by a  combined fit using the
single-lepton channel and the theoretical prediction at NLO is in agreement with
this measurement.  In addition differential cross sections with respect to
photon \pt and $\eta$ are measured.  These results also agree with the
theoretical predictions at NLO within uncertainties.

\section*{Acknowledgements}
% Acknowledgements for papers with collision data
% Version 6-Mar-2017

% Standard acknowledgements start here
%----------------------------------------------
We thank CERN for the very successful operation of the LHC, as well as the
support staff from our institutions without whom ATLAS could not be
operated efficiently.

We acknowledge the support of ANPCyT, Argentina; YerPhI, Armenia; ARC, Australia; BMWFW and FWF, Austria; ANAS, Azerbaijan; SSTC, Belarus; CNPq and FAPESP, Brazil; NSERC, NRC and CFI, Canada; CERN; CONICYT, Chile; CAS, MOST and NSFC, China; COLCIENCIAS, Colombia; MSMT CR, MPO CR and VSC CR, Czech Republic; DNRF and DNSRC, Denmark; IN2P3-CNRS, CEA-DSM/IRFU, France; SRNSF, Georgia; BMBF, HGF, and MPG, Germany; GSRT, Greece; RGC, Hong Kong SAR, China; ISF, I-CORE and Benoziyo Center, Israel; INFN, Italy; MEXT and JSPS, Japan; CNRST, Morocco; NWO, Netherlands; RCN, Norway; MNiSW and NCN, Poland; FCT, Portugal; MNE/IFA, Romania; MES of Russia and NRC KI, Russian Federation; JINR; MESTD, Serbia; MSSR, Slovakia; ARRS and MIZ\v{S}, Slovenia; DST/NRF, South Africa; MINECO, Spain; SRC and Wallenberg Foundation, Sweden; SERI, SNSF and Cantons of Bern and Geneva, Switzerland; MOST, Taiwan; TAEK, Turkey; STFC, United Kingdom; DOE and NSF, United States of America. In addition, individual groups and members have received support from BCKDF, the Canada Council, CANARIE, CRC, Compute Canada, FQRNT, and the Ontario Innovation Trust, Canada; EPLANET, ERC, ERDF, FP7, Horizon 2020 and Marie Sk{\l}odowska-Curie Actions, European Union; Investissements d'Avenir Labex and Idex, ANR, R{\'e}gion Auvergne and Fondation Partager le Savoir, France; DFG and AvH Foundation, Germany; Herakleitos, Thales and Aristeia programmes co-financed by EU-ESF and the Greek NSRF; BSF, GIF and Minerva, Israel; BRF, Norway; CERCA Programme Generalitat de Catalunya, Generalitat Valenciana, Spain; the Royal Society and Leverhulme Trust, United Kingdom.

The crucial computing support from all WLCG partners is acknowledged gratefully, in particular from CERN, the ATLAS Tier-1 facilities at TRIUMF (Canada), NDGF (Denmark, Norway, Sweden), CC-IN2P3 (France), KIT/GridKA (Germany), INFN-CNAF (Italy), NL-T1 (Netherlands), PIC (Spain), ASGC (Taiwan), RAL (UK) and BNL (USA), the Tier-2 facilities worldwide and large non-WLCG resource providers. Major contributors of computing resources are listed in Ref.~\cite{ATL-GEN-PUB-2016-002}.
%----------------------------------------------

\printbibliography
\newpage
% ATLAS Collaboration author list
% Data extracted on 27-Apr-2017 for paper reference TOPQ-2015-21
%\usepackage{a4wide}
\begin{flushleft}
{\Large The ATLAS Collaboration}

\bigskip

M.~Aaboud$^\textrm{\scriptsize 137d}$,
G.~Aad$^\textrm{\scriptsize 88}$,
B.~Abbott$^\textrm{\scriptsize 115}$,
O.~Abdinov$^\textrm{\scriptsize 12}$$^{,*}$,
B.~Abeloos$^\textrm{\scriptsize 119}$,
S.H.~Abidi$^\textrm{\scriptsize 161}$,
O.S.~AbouZeid$^\textrm{\scriptsize 139}$,
N.L.~Abraham$^\textrm{\scriptsize 151}$,
H.~Abramowicz$^\textrm{\scriptsize 155}$,
H.~Abreu$^\textrm{\scriptsize 154}$
% ...
\bigskip
\\
% ...
$^{12}$ Institute of Physics, Azerbaijan Academy of Sciences, Baku, Azerbaijan\\
% ...
$^{88}$ CPPM, Aix-Marseille Universit{\'e} and CNRS/IN2P3, Marseille, France\\
% ...
$^{115}$ Homer L. Dodge Department of Physics and Astronomy, University of Oklahoma, Norman OK, United States of America\\
% ...
$^{119}$ LAL, Univ. Paris-Sud, CNRS/IN2P3, Universit{\'e} Paris-Saclay, Orsay, France\\
% ...
$^{137}$ $^{(a)}$ Facult{\'e} des Sciences Ain Chock, R{\'e}seau Universitaire de Physique des Hautes Energies - Universit{\'e} Hassan II, Casablanca; $^{(b)}$ Centre National de l'Energie des Sciences Techniques Nucleaires, Rabat; $^{(c)}$ Facult{\'e} des Sciences Semlalia, Universit{\'e} Cadi Ayyad, LPHEA-Marrakech; $^{(d)}$ Facult{\'e} des Sciences, Universit{\'e} Mohamed Premier and LPTPM, Oujda; $^{(e)}$ Facult{\'e} des sciences, Universit{\'e} Mohammed V, Rabat, Morocco\\
% ...
$^{139}$ Santa Cruz Institute for Particle Physics, University of California Santa Cruz, Santa Cruz CA, United States of America\\
% ...
$^{151}$ Department of Physics and Astronomy, University of Sussex, Brighton, United Kingdom\\
% ...
$^{154}$ Department of Physics, Technion: Israel Institute of Technology, Haifa, Israel\\
$^{155}$ Raymond and Beverly Sackler School of Physics and Astronomy, Tel Aviv University, Tel Aviv, Israel\\
% ...
$^{161}$ Department of Physics, University of Toronto, Toronto ON, Canada\\
% ...
$^{a}$ Also at Department of Physics, King's College London, London, United Kingdom\\
$^{b}$ Also at Institute of Physics, Azerbaijan Academy of Sciences, Baku, Azerbaijan\\
$^{c}$ Also at Novosibirsk State University, Novosibirsk, Russia\\
$^{d}$ Also at TRIUMF, Vancouver BC, Canada\\
$^{e}$ Also at Department of Physics {\&} Astronomy, University of Louisville, Louisville, KY, United States of America\\
% ...
$^{*}$ Deceased
\end{flushleft}

% Created with xml2latex.py

%\clearpage
%\part*{Auxiliary material}
%\input{appendix}
\end{document}